\documentclass[12pt,preprint]{aastex}
\usepackage{amsmath}
\usepackage{color}

\begin{document}

\newcommand{\lya}{Ly$\alpha$}
\newcommand{\nv}{N\,{\footnotesize V}}
\newcommand{\siiv}{Si\,{\footnotesize IV}}
\newcommand{\civ}{C\,{\footnotesize IV}}
\newcommand{\ciii}{C\,{\footnotesize III}}
\newcommand{\mgii}{Mg\,{\footnotesize II}}
\newcommand{\hr}{H$\gamma$}
\newcommand{\hb}{H$\beta$}
\newcommand{\ha}{H$\alpha$}
\newcommand{\pr}{Pa$\gamma$}
\newcommand{\oii}{[O\,{\footnotesize II}]}
\newcommand{\oiii}{[O\,{\footnotesize III}]}
\newcommand{\hei}{He\,{\footnotesize I}}
\newcommand{\heii}{He\,{\footnotesize II}}
\newcommand{\nii}{[N\,{\footnotesize II}]}
\newcommand{\sii}{[S\,{\footnotesize II}]}
\newcommand{\feii}{Fe\,{\footnotesize II}}

\newcommand{\kmps}{$\rm km~s^{-1}$}
\newcommand{\mbh}{$M_{\rm BH}$}
\newcommand{\msun}{$M_{\odot}$}
\newcommand{\rielr}{$R_{\rm IELR}$}
\newcommand{\rsub}{$R_{\rm sub}$}

\title{Detection of the Intermediate-width Emission Line Region in Quasar OI 287 with the Broad Emission Line Region Obscured by the Dusty Torus}
\author{Zhenzhen Li\altaffilmark{1,2,3}, Hongyan Zhou\altaffilmark{2,1,3}, Lei Hao\altaffilmark{4}, Huiyuan Wang\altaffilmark{1,3}, Tuo Ji\altaffilmark{2,3}, Xiheng Shi\altaffilmark{2}, Bo Liu\altaffilmark{1,2,3}, Shaohua Zhang\altaffilmark{2}, Wen-Juan Liu\altaffilmark{1,2,3}, Xiang Pan\altaffilmark{1,2,3}, Peng Jiang\altaffilmark{1,3}}
\altaffiltext{1}{Department of Astronomy, University of Science and Technology of China, Hefei, Anhui 230026, China; lizz08@mail.ustc.edu.cn}
\altaffiltext{2}{Polar Research Institute of China, Jinqiao Rd. 451, Shanghai, 200136, China; zhouhongyan@pric.org.cn}
\altaffiltext{3}{Key Laboratory for Research in Galaxies and Cosmology, University of Science and Technology of China, Chinese Academy of Sciences, Hefei, Anhui 230026, China}
\altaffiltext{4}{Key Laboratory for Research in Galaxies and Cosmology, Shanghai Astronomical Observatory, Chinese Academy of Sciences, 80 Nandan Road, Shanghai 200030, China; haol@shao.ac.cn}
\begin{abstract}
The existence of intermediate-width emission line regions (IELRs) in active galactic nuclei has been discussed for over two decades. A consensus, however, is yet to be arrived at due to the lack of convincing evidence for their detection. We present a detailed analysis of the broadband spectrophotometry of the partially obscured quasar OI 287. The ultraviolet intermediate-width emission lines (IELs) are very prominent, in high contrast to the corresponding broad emission lines (BELs) which are heavily suppressed by dust reddening. Assuming that the IELR is virialized, we estimated its distance to the central black hole of $\sim 2.9$ pc, similar to the dust sublimation radius of $\sim 1.3$ pc. Photo-ionization calculations suggest that the IELR has a hydrogen density of $\sim 10^{8.8}-10^{9.4} ~ \rm cm^{-3}$, within the range of values quoted for the dusty torus near the sublimation radius. Both its inferred location and physical conditions suggest that the IELR originates from the inner surface of the dusty torus. In the spectrum of this quasar, we identified only one narrow absorption-line system associated with the dusty material. With the aid of photo-ionization model calculations, we found that the obscuring material might originate from an outer region of the dusty torus. We speculate that the dusty torus, which is exposed to the central ionizing source, may produce IELs through photo-ionization processes, while also obscure BELs as a natural ``coronagraph''. Such a ``coronagraph'' could be found in a large number of partially obscured quasars and be a useful tool to study IELRs.
\end{abstract}


\section{INTRODUCTION}
Emission lines are one of the most important features in the spectra of active galactic nuclei (AGNs). Line velocities, profiles, and intensities can provide us with an opportunity to understand the geometry, kinematics, and physical conditions of the emitting gas around the AGNs. Traditionally, AGN emission lines are often categorized into two groups according to their line widths: broad emission lines (BELs) with full width at half maximum (FWHM) $\sim$ 5000 \kmps\ and narrow emission lines (NELs) with FWHM $\sim$ 500 \kmps. Permitted and semi-forbidden lines are present in both BELs and NELs, while forbidden lines only in NELs. Variability is often observed in BELs, but rarely in NELs. Such a split is commonly interpreted as that BELs and NELs originate from two distinct regions: broad emission line regions (BELRs) have a larger velocity dispersion and a higher electron density ($n_{\rm e} \sim 10^9-10^{13} ~ \rm cm^{-3}$) located in a compact region within $\sim$ 1 pc from the central super-massive black hole; narrow emission line regions (NELRs) are much more extended, $\sim$ 0.1--1 kpc in size, and have a smaller velocity dispersion and a lower electron density ($n_{\rm e} \sim 10^{3}-10^{6} ~ \rm cm^{-3}$).

Such a crude division of emission-line regions into BELRs and NELRs may be oversimplified. Some researchers suggested that AGNs also contain intermediate-width emission line regions (IELRs) producing intermediate-width emission lines (IELs) with FWHM $\sim$ 2000 \kmps\ (e.g., Wills et al. 1993; Brotherton et al. 1994a, b, 1996; Mason et al. 1996; Sulentic et al. 2000; Hu et al. 2008; Zhu et al. 2009; Zhang 2011, 2013). However, the presumed IELR, involved in many debates, has not been widely accepted since proposed in the 1990s. For instance, the IELR is considered as an outer part of the traditional BELR in some studies (e.g., Brothertton et al. 1994, Sulentic 2000, Hu et al, 2008, Zhu et al. 2009, Zhang et al. 2011, 2013), while as an inner part of the NELR in others (e.g., Sulentic 1999). Brotherton et al. (1994) concluded that the redshift of the IELR is consistent with the systemic redshift, whereas Hu et al. (2008) claimed that the IELR is systematically redshifted in the rest frame of AGNs and arises from inflowing gas. These conflicts about the IELR may be due to the uncertainties of line decomposition, using the methods of emission-line fit (Mason et al. 1996; Hu et al. 2008; Zhu et al. 2009) or the principal components analysis (PCA) technique (Brotherton et al. 1994 b; Zhang 2011).

The IELs in normal quasars are suggested to be very weak compared with the BELs and NELs, because dust mixed in the IELR absorbs most of the ionizing photons and thus suppresses the line emission (Netzer \& Lar, 1993). This weakness results in the difficulty of detecting the IELs. Partially obscured quasars may provide an opportunity to reliably detect the IELs from the quasar emission-line spectra. In these quasars, the emission from their central accretion disk and BELR may be significantly obscured by dust. If the IELR does exist, the IELs would become prominent in the shorter wavelength range where the BELs are heavily suppressed. In this paper, we report such a quasar -- OI 287 -- found from a compiled sample of \textit{Hubble Space Telescope} (\textit{HST}) Faint Object Spectrograph (FOS) spectra, which was constructed to systematically study the IELR (Li et al. 2015, in preparation). OI 287 is very remarkable in the sample due to its prominent IELs of \lya\ and \civ, which may present direct evidence of an IELR and provide a robust avenue to understand the properties of IELRs.

This paper is organized as follows. In Section 2, we describe the observations and data reduction; in Section 3, we analyze the observational data including emission lines, broadband spectral energy distributions (SED) and absorption lines; in Section 4, we discuss the properties of IELRs and the obscuring material and find more similar objects; finally, we give a brief summary in Section 5. Throughout this paper, we use the cosmological parameters $H_0 = 70 ~ \rm km~s^{-1} \rm Mpc^{-1}$, $\Omega_{\rm M} = 0.3$, and $\Omega_{\Lambda} = 0.7$.
\section{OBSERVATIONS AND DATA REDUCTION}
The ultraviolet (UV) spectrum of OI 287 was obtained by FOS on board \textit{HST} using the G190H grating on May 21, 1992. The \textit{HST} spectrum covers a wavelength range from 1590 \AA\ to 2310 \AA\ with a spectral resolution of $R \sim$ 1300. The fully-processed and well-calibrated \textit{HST}/FOS spectra were retrieved from the \textit{HST}/FOS Spectral Atlas\footnote{http://hea-www.harvard.edu/\~{}pgreen/HRCULES.html} compiled by Kuraszkiewicz et al. (2004).

OI 287 was also spectroscopically observed by the Sloan Digital Sky Survey (SDSS, York et al. 2000) on October 31, 2002. The spectrum provides a wavelength range of $\lambda \sim$ 3800--9200 \AA\ with a spectral resolution of $R \sim$ 2000. We extracted the spectrum from SDSS data release 7 (DR7; Abazajian et al. 2009).

To acquire the near-infrared (NIR) spectrum, we performed follow-up spectroscopic observations of OI 287 using TripleSpec (Wilson et al. 2004) on the 200--inch Hale Telescope at Palomar Observatory on February 23, 2013. A slit of $1.1''$ was chosen to match the seeing and four 180 s exposures were taken in an A-B-B-A dithering mode with the primary configuration of the instrument. This gave a spectral resolution of $R \sim$ 3500 and a wavelength coverage of $\lambda \sim$ 0.97--2.46 $\mu$m. Two telluric standard stars were observed quasi-simultaneously for flux calibration. The data were reduced with the \textit{Triplespectool} package, a modified version of \textit{Spextool} (Cushing et al. 2004).

To narrow down the wavelength gaps, we also performed spectroscopic observations of OI 287 using DoubleSpec\footnote{http://www.astro.caltech.edu/palomar/200inch/dbl\_{}spec/dbsphome.html} mounted on the Hale Telescope on April 22, 2014. A $2''$ slit was chosen to match the seeing and two 600 s exposures were taken using the 600 lines mm$^{-1}$ gratings, one blazed at 3780 \AA\ and the other at 9500 \AA. These settings yield two wavelength coverages of $\lambda \sim$ 3150--5850 \AA\ and $\lambda \sim$ 7840--10700 \AA, respectively. The BD+75D325 standard star was observed quasi-simultaneously for flux calibration. Wavelength calibration was carried out using an Fe/Ar lamp for the blue portion and He/Ne/Ar lamp for the red portion. The data reduction was accomplished with standard procedures using IRAF\footnote{IRAF is distributed by the National Optical Astronomy Observatory, which is operated by the Association of Universities for Research in Astronomy, Inc., under cooperative agreement with the National Science Foundation.}. We combined the DoubleSpec and SDSS spectrum to form one spectrum covering a wavelength range of $\lambda \sim$ 3200--10700 \AA.

The spectroscopic observations are summarized in Table 1. We also collected broadband photometric data of OI 287 from available large sky surveys, from UV (the \textit{Galaxy Evolution Explorer} or \textit{GALEX}, Morrissey et al. 2007), through optical (SDSS) to infrared (the UKIRT Infrared Deep Sky Survey or UKIDSS, Lawrence et al. 2007; the Wide-field Infrared Survey Explorer or WISE, Wright et al. 2010). The details of the multi-wavelength photometric data are presented in Table 2. All of the spectroscopic and photometric data have been corrected for a Galactic reddening of $E(B-V)$=0.059 using the updated dust map of Schlafly \& Finkbeiner (2011) and converted to the rest frame of the quasar using the redshift $z=0.4443$ (Schneider et al. 2010) before conducting analysis.
\section{DATA ANALYSIS and RESULTS}
\subsection{Emission Line Spectrum}
The observed spectra are displayed in Figure 1. It is striking that the UV emission lines of \lya\ and \civ\ are much narrower compared with the optical and NIR broad lines. To make a clear comparison, we show the profiles of \lya\ and \ha\ in their common velocity space as an example in the insert panel. In order to further study the unusual emission lines, we first subtract the underling continuum from the observed spectra by fitting the spectra in three spectral ranges. (1) We fit the continuum of the \textit{HST} spectrum using a single power law in continuum windows free from strong emission lines. The derived continuum, with a spectral index $\alpha \approx -1.75$ ($f_{\nu} \propto \nu^{\alpha})$, is redder compared with the quasar composite spectrum ($\alpha = -0.46$, Vanden Berk et al. 2001). (2) The continuum of the SDSS+DoubleSpec spectrum is modelled with the combination of a power law, a Balmer continuum (supplemented with blended high-order Balmer emission lines), and \feii\ multiplets, as described in detail in Dong et al. (2008). This continuum, with a spectral index of $\alpha \approx -2.04$, is also redder than the quasar composite spectrum. (3) The continuum of the TripleSpec spectrum is fitted with a second-order polynomial. We overplot these continuum models in Figure 1.

After subtracting the continuum model, we obtain the emission line spectrum. Figure 2 displays the strong permitted emission lines, including \lya, \civ, \mgii, \hr, \hb, \ha, and \hei, in their common velocity space. \lya\ is dominated by an intermediate-width component with FWHM $\approx$ 2000 \kmps, while its broad component is almost completely absent. Similarly, the adjacent line \civ\ also presents a prominent intermediate-width component, but has a more apparent broad component. At longer wavelengths, the emission lines such as \mgii, \hr, \hb, \ha, and \hei\ are dominated by their broad components. The trend that broad components gradually become weaker toward shorter wavelengths indicates that the BELR of OI 287 may be reddened.

To quantify the different components of emission lines, we decompose these emission lines into three components: broad, narrow, and intermediate-width. Each component is modelled with a single Gaussian. The same components in different lines are assumed to have the same redshift and line width. For the doublets of \civ\ and \mgii, each doublet component is modelled separately with their relative intensity ratios fixed at $1:1$, assuming that the emission is optically thick. \hb\ shows a red wing extending underneath the \oiii\ double lines (the ``red shelf'', Meyers \& Peterson 1985; V{\'e}ron et al. 2002), which might be attributed to (a) broad \hb, (b) broad \oiii\ $\lambda\lambda$4959, 5007, (c) \hei\ $\lambda\lambda$4922, 5016, or (d) \feii. This feature unlikely originates from \hb\ emission since it is absent in \hr\ and \ha. Detailed study of this feature is beyond the scope of this paper and we use an additional broad Gaussian to eliminate its influence. The red wing of \hei\ is blended with \pr, which is also modelled by the three components with the similar profiles as the corresponding components of Balmer lines. The UV lines of \nv\ $\lambda$1240 and \siiv\ $\lambda$1397, usually obvious in the spectrum of AGNs, are very weak in this spectrum. These weak lines may also provide important information and we also fit them using the three components to detect their upper limits. The forbidden lines including \oiii\ $\lambda\lambda$4363, 4959, 5007, \nii\ $\lambda\lambda$6548, 6583, and \sii\ $\lambda\lambda$6716, 6731 are also fitted. Most of these lines are fitted using one narrow component, except for the three \oiii\ lines, which are fitted with two components, one narrow component for the line core and one free Gaussian component for the blue wing. The relative intensity ratios of \oiii\ $\lambda\lambda$4959, 5007 (for both core and wing components) and \nii\ $\lambda\lambda$6548, 6583 are fixed to their theoretical value of $1:3$. Then we simultaneously fit all of these emission lines using an Interactive Data Language (IDL) code based on MPFIT (Markwardt 2009), which performs $\chi^2$--minimization by the Levenberg – Marquardt technique. During the fitting process, absorption lines in \civ, \mgii, and \hei\ are carefully masked. The best-fit results are shown in Figure 2 and the emission-line parameters are summarized in Table 3 (Model 1).

In order to check whether the fitting results are model dependent, we decompose these lines also with a different method. Because the three components dominate different lines (for example, NELs are dominant in the forbidden lines, IELs are dominant in the UV permitted lines, and BELs are dominant in the optical permitted lines), their shifts and profiles can be reliably obtained in the corresponding lines. We first obtain the redshift and line width of the NELs from the core of the \oiii\ double lines, the BELs from the \ha\ emission line, and the IELs from the \lya\ emission line. Then, we use the three-component model to fit all of the emission lines. These best-fit results, also shown in Table 3 (Model 2), are similar to those from the  method above, indicating the line decompositions are independent of models we used.

Due to the weakness of the UV BELs in the quasar OI 287, the UV IELs become prominent and thus can be reliably measured. We made a simulation by measuring the \lya\ IEL as an example to investigate the dependence of measurement accuracy over the IEL/BEL flux ratios. The simulation result shows that the 1-$\sigma$ measurement flux errors for OI 287 are greatly reduced from 18.7\% to 0.8\%. Meanwhile, the profile decomposition uncertainties for the optical/NIR IELs can also be reduced, because their redshifts and profiles can be fixed to those of the prominent UV IELs. We also carried out a simulation by measuring the \ha\ IEL as an example to inspect this fitting strategy. This simulation shows that by fixing the redshift and profile of the \ha\ IEL to those of \lya\ IEL, the 1-$\sigma$ measurement flux errors are reduced from 37.3\% to 15.4\%. (See Appendix A for how we perform the simulations and calculate the measurement errors.)

With the measurements of emission lines, we first investigate the extinction using the Balmer decrement (\ha/\hb). The intrinsic Balmer decrement under the CASE B condition is \ha/\hb\ = 2.73--3.30 (Osterbrock 2006). The measured Balmer decrement for NELs of OI 287, \ha/\hb\ $=3.06\pm0.19$, is within the theoretical range. Similarly, the Balmer decrement for IELs, \ha/\hb\ = 2.74$\pm0.90$, is also consistent with the theoretical value. These results indicate that the NELR and IELR are not reddened. On the other hand, the Balmer decrement for BELs, \ha/\hb\ = 5.79$\pm0.07$, is much larger than the theoretical value, indicating that the BELR is obviously reddened. We further investigate the BELR extinction through intensity ratios of BELs in OI 287 to BELs in the quasar composite. As a rough approximation, we evaluate the BELs intensities of the composite quasar spectrum by simply employing the measurements for whole emission lines (\lya, \civ, \mgii, \hr, \hb, and \ha\ from Vanden Berk et al. (2001); \hei\ and \pr\ from Glikman et al. (2006)), since these emission lines in the composite spectrum are dominated by their broad component. \hei\ is blended with \pr, therefore we use the summed intensity (\hei\ + \pr) for both OI 287 and the composite spectrum. Figure 3 shows the derived intensity ratios of BELs in OI 287 to BELs in the quasar composite, which is normalized to unity at \hei\ + \pr. The intensity ratios gradually decrease toward shorter wavelengths, suggesting that the BELs of OI 287 are reddened by dust. We fit the intensity ratios using three commonly used extinction curves of the Small Magellanic Cloud (SMC), Large Magellanic Cloud and Milky Way. The parametrizations of these extinction curves are taken from Pei (1992). All of these three extinction curves can interpret the BEL intensity ratios, but the SMC-like extinction, which yields an $E(B-V)$ of 0.29$\pm0.03$, is relatively more likely. Anyhow, all of these extinction curves indicate that the BELR of OI 287 is obscured by dust.

Since dust extinction is stronger toward shorter wavelengths, the broad components of \lya\ and \civ\ are heavily suppressed, which can naturally explain why the intermediate-width component in these two lines are prominent. On the other hand, the prominent IELs in the UV imply that they should arise from a distinct region that is not obviously obscured by dust. The location of the IELR will be further discussed in Section 4.1.
\subsection{Broadband Spectral Energy Distribution}
Since the BELR of OI 287 is obscured, the central accretion disk is likely to be also obscured because the optical-emitting part of the accretion disk is smaller than the BELR. Therefore, we investigate the SED, which can reveal the properties of the accretion disk, to confirm the extinction.

With the multi-wavelength spectroscopic and photometric data, we construct the broadband SED of OI 287 spanning from 1100 \AA\ to 15 $\mu$m in the rest frame. Figure 4 displays the broadband SED. From the infrared to optical band, photometric and spectroscopic data are well consistent with each other, implying that the variation is not significant among different observation epochs. The long-term intensive monitoring observations by the Catalina Sky Survey\footnote{http://nesssi.cacr.caltech.edu/DataRelease/} for nearly ten years (2005/04/10--2014/01/23) demonstrate that the optical \textit{V}-band variation amplitudes of OI 287 are within 0.1 magnitude. The only exception is that the \textit{GALEX} photometry is larger than the \textit{HST} spectrum for about 2 times. It is not clear whether this difference is caused by the calibration uncertainties or intrinsic variabilities. As the \textit{HST} spectrum provides more information of the continuum, we use the \textit{HST} data in the following analysis.

For comparison, a rescaled quasar composite spectrum (Zhou et al. 2010), which is obtained by combining the SDSS composite ($\lambda <$ 3000 \AA; Vanden Berk et al. 2001), the NIR template (3000 \AA\ $< \lambda <$ 2 $\mu$m; Glikman et al. 2006), and mid-infrared (MIR) template ($\lambda >$ 2 $\mu$m; Netzer et al. 2007), is overplotted in Figure 4 and normalized to the SED of OI 287 at the $W3$ band. The observed SED of OI 287 is nearly identical to the composite spectrum in the longer wavelength range ($\lambda > 7000$ \AA), while gradually deviates from the composite spectrum toward the shorter wavelength range ($\lambda < 7000$ \AA). This indicates that the accretion disk may be reddened.

To further analyze the reddening of the accretion disk emission, we model the broadband SED using the following four components. (1) \textbf{Reddened power law:}  We use a power law with a spectral index $\alpha$ reddened by the SMC extinction curve with an $E(B-V)$ to represent the continuum radiation from the obscured accretion disk. (2) \textbf{Scattered power law:} Spectropolarimetrice observations showed that OI 287 has significant polarization of up to 8\% caused by scattering (Moore and Stockman 1981; Goodrich and Miller 1988; Rudy and Schmidt 1988). In the observer-frame wavelength range of $\lambda \approx$ 4600--8000 \AA\ (3200--5500 \AA\ in the quasar rest frame), the polarization modestly increases toward shorter wavelength (Rudy and Schmidt 1988), inferring that the scattering may be more important in the UV. The scattered light, arising from an extensive region, should not be reddened as indicated by the unreddened NELR. Based on these considerations, we add an unreddened power law with the same spectral index $\alpha$ to represent the scattered component. (3) \textbf{Hot black body:} The near infrared bump is caused by the thermal radiation of hot dust in the inner region of the dusty torus. We use a black body with temperature {$T_{\rm HD}$ in the range of 500--1500 K to model the hot dust radiation. (4) \textbf{Warm black body:} From middle to far infrared, the SED is dominated by the radiation from warm dust in the outer region of the dusty torus. A black body with temperature $T_{\rm CD} <$ 500 K is used to model the warm dust radiation. We do not consider the contribution of the quasar host galaxy. There is no significant feature of starlight found in the spectrum. In addition, from the SDSS \textit{r} band photometry data, we do not find significant difference between the point spread function (PSF) magnitude (17.71 mag) and the model magnitude (17.66 mag), which also implies that the contribution of starlight is not important.

The best-fit results are displayed in Figure 4. The four-component model provides a good fit to the observational data and the derived parameters are also reasonable. The derived spectral index of the power law is $\alpha=-0.56\pm0.13$, similar to that of the quasar composite spectrum $-0.46$. The $E(B-V)$ derived from the SED fitting is $0.31\pm0.03$, which agrees well with that derived from the BELs ($0.29\pm0.03$) in Section 3.1. The fraction of scattered component increases from 5.3\% at 5500 \AA\ to 10.1\% at 3200 \AA, roughly consistent with the observed degree of polarization from p $\approx$ 6\% to 8\% across the wavelength range (Rudy and Schmidit 1988). Our fitting suggests that the hot dust has a temperature of $T_{\rm HD}$ = 1250$\pm60$ K, similar to the value found by Glikman et al. (2006) by modelling the composite quasar spectrum. By combining the measured hot dust luminosity ($L_{\rm HD}$) with the bolometric luminosity $L_{\rm bol}$ evaluated using the bolometric correction $L_{\rm bol}=9 \lambda L_{\lambda}$(5100 \AA) (Kaspi et al. 2000), we derive an estimation for the covering factor $f_{\rm C}$ of the hot dust as $L_{\rm HD}/L_{\rm bol} \approx 7\%$. We will use this estimation to study the properties of the IELR in Section 4.1.
\subsection{Absorption Line Spectrum}
The results above indicate that the accretion disk and BELR of OI 287 are obscured by dust located somewhere along the line-of-sight. Since the dust is always mixed with gas, we can trace the location of the dust via the gas absorption lines. In the spectrum of OI 287, we found one absorption-line system, involving \civ\ $\lambda\lambda$1548, 1551, \mgii\ $\lambda\lambda$2796, 2803, and \hei$^*$ $\lambda\lambda$3189, 3889, 10830. We show the local spectrum around these absorption lines in the left column of Figure 5.

We normalize the absorption lines using the best-fit models above of emission lines (NEL, IEL and BEL) and SED (reddened power law, scattered power law, hot black body and warm black body). First, we subtract the models of NEL, IEL, black body, and the scattered power law from the observed spectrum, assuming that these emission regions are not covered by the absorption gas. This is based on the following considerations. (a) These regions are located in extensive regions far away from the compact central source, which are usually hard to obscure. (b) As shown in Section 3.1, NELs and IELs are not significantly dust reddened, which implies that neither one is covered by the obscuring gas. (c) As shown in Figure 5, after subtracting the emission models above, there is almost no residual flux at the bottom of the \civ\ and \hei\ absorption lines. It can be readily interpreted as that the absorption gas fully covers the BELR and accretion disk (similar to the dust mentioned above) and does not obscure the light from the other regions, i.e. IELR, hot and warm dust, and the scattering region. We then normalize the observed spectrum by the sum of the power law and BELs. These two emission regions are assumed to be covered by gas, as indicated by their reddening by dust.

The normalized absorption lines are shown in the right column of Figure 5. Most of these absorption lines have similar shifts and line widths, except those of the \civ\ doublet, which have larger blueshifts, broader line widths, and multiple absorption troughs. Therefore, we cannot acquire reliable results from the \civ\ absorption lines and focus on the other absorption lines in the following analysis. We simultaneously fit all of the absorption lines using a single Gaussian for each line, assuming that they have the same shifts and line widths. The best-fit results are presented in the right column of Figure 5. The equivalent width ratio of the \mgii\ doublets, EW(\mgii\ $\lambda$2796)/EW(\mgii\ $\lambda$2803) = 1.84$\pm0.21$, is consistent with the theoretical ratio of 2.00 for unsaturated lines with a full coverage. The \hei$^*$ $\lambda$3189 absorption line is marginally detected with a 3-$\sigma$ upper limit of EW(\hei$^*$ $\lambda$3189) $\lesssim$ 0.21 \AA. The ratio of EW(\hei$^*$ $\lambda$3889)/EW(\hei$^*$ $\lambda$3189) $\gtrsim$ 3.04 is also consistent with the theoretical ratio of 3.08. These results suggest that the absorption gas fully covers the accretion disk and BELR, and that the absorption lines are not saturated. Under this scenario, the column density of corresponding ions can be derived from their equivalent widths using the equation from Jenkins (1986), $N=(m_e c^2/\pi e^2 f \lambda^2)~\rm EW$, where $f$ is the oscillator strength, $m_{\rm e}$ is the electron mass and $e$ is the electron charge. We derive the column densities for Mg$^+$ from the \mgii\ $\lambda$2796 line with $N_{\rm Mg^+}=(1.91\pm0.54) \times 10^{13} ~ \rm cm^{-2}$, and for \hei$^*$ from the \hei$^*$ $\lambda$3889 line with $N_{\rm HeI^*}=(1.93\pm0.22) \times 10^{14} ~ \rm cm^{-2}$. In addition, with the same redshift and line width, we also measure the 3-$\sigma$ upper limit of the \ha\ absorption line, EW(\ha) $\lesssim$ 0.09 \AA, and acquire the maximum column density of hydrogen at the n = 2 level, $N_{\rm HI,2} \lesssim 3.6 \times 10^{11} ~\rm cm^{-2}$. These measurements will be used to study the properties of the absorption gas in Section 4.2.
\section{DISCUSSION}
\subsection{Properties of the Intermediate-width Emission Line Region}
With the measurements of the \hb\ broad line width and extinction corrected continuum luminosity at 5100 \AA, the central black hole mass (\mbh) is estimated to be $1.46^{+1.89}_{-0.82} \times 10^9$ \msun\footnote{According to the definition, one should use the line width of BEL+IEL \hb\ (i.e., the NEL subtracted \hb) to estimate \mbh. This is the standard procedure in normal quasars and gives an estimate of $M_{\rm BH} = 1.32^{+1.60}_{-0.72} \times 10^9$ \msun. However, in OI 287 the BEL of \hb\ is attenuated and thus the IEL is more prominent than normal quasars. Therefore we only use the \hb\ BEL to estimate the \mbh. The uncertainty between these two extremes will be propagated to all the following properties derived from the \mbh.} by employing the empirical formula of Wang et al. (2009). By combining \mbh\ with the measurement of FWHM(IELs), and assuming that the clouds in the IELR are virialized, the distance of the IELR to the central black hole (\rielr) can be derived as $R_{\rm IELR} = G M_{\rm BH}/(f \rm {FWHM(IELs)})^2$, where $G$ is the gravitational constant and $f$ is a scaling factor. With a simple approximation of an isotropic IELR and Gaussian-profile IELs, $f=\sqrt{3}/2.354$\footnote{For an isotropic IELR, the velocity dispersion ($\sigma)$ along the line-of-sight ($\sigma_{\rm line}$) is equal in all directions, $\sigma=\sqrt{3} \sigma_{\rm line}$. For a Gaussian profile of IELs, $\sigma_{\rm line}=\rm {FWHM(IELs)}/2.354$. Thus, the scale factor is $f \equiv \sigma/\rm{FWHM(IELs)} = \sqrt{3}/2.354$.}. With these assumptions, we derive an estimate of $R_{\rm IELR}=2.9^{+4.3}_{-2.1}$ pc. This is similar to the dust sublimation radius (\rsub) of 1.3 pc, estimated using the formula in Barvainis 1987, $R_{\rm sub} = 1.3 ~ L_{uv,46}^{0.5} ~ T_{1500}^{-2.8} ~ \rm pc$, where $L_{uv,46}$ is the UV luminosity of the central source in units of $10^{46} ~ \rm erg~ s^{-1}$, and $T_{1500}$ is the grain sublimation temperature in units of 1500 K. The coincidence of these two values implies that the IELR may be located in the inner part of the dusty torus. According to the widely accepted unified model of AGNs (e.g., Antonucci 1993), gas on the inner surface of the dusty torus is exposed to the central ionizing source. As a result, it can be inferred that illuminated gas in this region may produce emission lines through photo-ionization processes.

If the IELs are produced through photo-ionization processes, we can constrain the IELR physical conditions by comparing the observed IELs with those of the photo-ionization models. We perform a simulation using CLOUDY (Version 13.03, Ferland et al. 1998) by considering a gas slab, which is illuminated by a quasar with an SED defined by Mathews \& Ferland (1987, hereafter MF87). The quasar monochromatic luminosity at 1450 \AA\ is scaled to that of OI 287 after extinction correction, $\lambda L_{\lambda}$(1450 \AA) $\approx 5.4 \times 10^{45}~\rm erg~s^{-1}$, and the distance of the gas to the quasar (d) is fixed to the $R_{\rm IELR}$ of OI 287 (i.e., the hydrogen ionizing photon volume density is fixed to be $\Phi(\rm H) \approx 10^{17.2}~\rm photons~s^{-1}~cm^{-2}$). The covering factor is set to that of the hot dust ($\sim$ 7\% in Section 3.2), assuming that the IELR is located in the inner part of the dusty torus. The effect of dust grains is taken into account in the calculation. As the extinction of BELs in OI 287 can be described using the SMC-like extinction law, we model the dust with the same composition, consisting of graphite and silicate, and the same size distribution as SMC suggested in Weingartner \& Draine (2001). The total abundance of the gas and dust is assumed solar and the dust-to-gas ratio is set to that of the dusty torus of OI 287 estimated by the absorption lines (Section 4.2). We calculate a grid of models by varying the hydrogen density ($n_{\rm H}$) from $10^{3}$ to $10^{11}~\rm cm^{-3}$ and hydrogen column density ($N_{\rm H}$) from $10^{19}$ to $10^{23} ~\rm cm^{-2}$.

The calculated results are shown in Figure 6, where we plot the contours of EW(\lya), EW(\civ) and EW(\ha) as functions of $n_{\rm H}$ and $N_{\rm H}$. In each panel, the dashed lines denote the basic models and the filled areas represent the observed\footnote{Observed EWs are calculated as the ratio of the IEL flux to the extinction corrected continuum.} range with 1-$\sigma$ confidence level. Within the overlapping region, the parameters are constrained in a narrow range of $n_{\rm H} \sim 10^{8.8}-10^{9.4} ~\rm cm^{-3}$ and $N_{\rm H} \sim 10^{19.6}-10^{20.2} ~\rm cm^{-2}$. The derived $n_{\rm H}$ is consistent with the gas density near the sublimation radius suggested by the recent observation by Kishimoto et al. (2013) and modelling by Stern et al. (2014).

Adopting the derived $n_{\rm H}$ and $N_{\rm H}$, we predict the equivalent widths of all permitted IELs. The results are shown in Figure 7. The corresponding observed values are also plotted for comparison. Since the observed \nv\ and \siiv\ are marginally detected, we show their 3-$\sigma$ upper limit. The equivalent widths predicted by the photo-ionization model agree with their observed values within the uncertainties, which supports that IELs may be produced through photo-ionization processes. More importantly, adopting the covering factor estimated from hot dust emission gives results which are consistent with observations. Such results strongly favor that the IELR is closely associated with the hot dust region, the inner part of the dusty torus.

With the derived parameters above, the mass of emitting clouds in the IELR is estimated to be $M_{\rm IELR} = m_{\rm p}~4 \pi R_{\rm IELR}^2 f_{\rm C}~N_{\rm H} \approx 4.5 ~ M_{\odot}$, where $m_{\rm p}$ is the mass of a proton. Compared with the typical mass range of the dusty torus $\sim 10^{5}-10^{7} M_{\odot}$ (Mor et al. 2009), the mass of the IELR is negligible. This is generally consistent with the scenario that the IELR located in the inner part of the dusty torus. In addition, the thickness of the IELR is estimated to be $l \sim N_{\rm H}/n_{\rm H} = 2 \times 10^{-8}$ pc, which is also much smaller than the typical size of a dusty torus. However, clouds in the IELR may be not smoothly distributed. The dusty torus is suggested to be very clumpy or filamentary (e.g., Krolik \& Begelman 1988), which requires a small volume filling factor $f_{\rm F} \ll 1$ (Nenkova et al. 2002). If so, the IELR may consist of a large number of ionized clouds that are distributed throughout much larger region of the torus.
\subsection{Properties of the Obscuring Material}
As shown previously, the central accretion disk and BELR of OI 287 are obscured by dust, while the IELR and NELR are not. This difference implies that the obscuring material is likely associated with the dusty torus. More precisely, the properties of the obscuring material can be estimated using the gas absorption lines, since dust is always mixed with gas and only one set of gas absorption-line system is found in the spectrum of OI 287. To constrain the obscuring material, we also perform a simulation with CLOUDY by considering a dusty gas with SMC-type grains, which is illuminated by a quasar with an MF87 SED. The total abundance of the gas and dust is assumed to be solar and the dust-to-gas ratio of $A_{\rm v}/N_{\rm H}$ increases with a grid of $1.0, 1.5, 2.0, 2.5, 3.0 \times 10^{-22}~\rm cm^{2}$ \footnote{We do not consider larger dust-to-gas ratios, because almost all of the Mg is depleted into the dust grains for $A_{V}/N_{\rm H} = 3.0 \times 10^{-22}~\rm cm^{2}$}. For each dust-to-gas ratio, we calculate a 2-dimensional grid with variable $U$ of $10^{-1.5}-10^{-0.5}$ and $n_{\rm H}$ of $10^{3}-10^{8}~\rm cm^{-3}$ and stop the calculation when $A_{\rm v}$ reaches the observed value of 0.83 (With $R_{V}=2.87$ for SMC (Gordon \& Clayton 1998) and $E(B-V)=0.29$, we get $A_{V} \equiv R_{V} E(B-V)=0.83$). Figure 8 shows the calculated results. We use the observed ranges of $N_{\rm Mg^+}$, $N_{\rm HeI^*}$, and $N_{\rm HI,2}$ to constrain the parameters. When $A_{V}/N_{\rm H} = 3 \times 10^{-22}~\rm cm^{2}$, $N_{\rm Mg^+}$, $N_{\rm HeI^*}$ and $N_{\rm HI,2}$ form an overlap of $U \sim 10^{-0.9}-10^{-0.8}$ and $n_{\rm H} \sim 10^{4.0}-10^{6.5}~\rm cm^{-3}$. With $U$ and $n_{\rm H}$, the distance of absorber to central ionizing source is derived as $R_{\rm absorber}=(Q({\rm H}) / 4\pi c U n_{\rm H})^{0.5}$, where $Q(\rm H)$ is the number of ionizing photons, $Q({\rm H})=\int_{\nu}^{\infty} L_{\nu}/h \nu d \nu \approx 2.2 \times 10^{56}~\rm photons~s^{-1}$. In each panel of Figure 8, we also show the contour of $R_{\rm absorber}$ as functions of $U$ and $n_{\rm H}$. As shown in Panel (e) of Figure 8, in the overlapping region, $R_{\rm absorber}$ is constrained in the range of $\sim 10-200$ pc. This indicates that the accretion disk and BELR of OI 287 is obscured by dust located in an outer region of the dusty torus.

In brief, the derived properties of the obscuring material and IELR suggest that both of them are part of the dusty torus. The IELR, located in the inner part of the dusty torus, is exposed to the central ionizing source and produces IELs through photo-ionization processes; while the obscuring material, in an outer region of the dusty torus, obscures the central accretion disk and BELR like a ``coronagraph''. Figure 9 displays the cartoon for detecting the IELR in which the dusty torus can be treated as a ``coronagraph''. The central accretion disk and BELR are obscured by the boundary of the dusty torus, which can account for the observed facts of a reddened SED and BELs. However, the IELR is not fully obscured by the dusty torus, which yields the observed IELs.

In addition, as shown in the cartoon, a fraction of the IELR may also be obscured, especially the near side of the IELR (marked with ``B'' in Figure 9). Therefore, the observed intensities of the IELs might be smaller than their intrinsic ones. As a result, some parameters (such as $n_{\rm H}$, $N_{\rm H}$, and $M_{\rm IELR}$) derived from the strength of IELs could be moderately underestimated, but not change dramatically. The qualitative properties of the IELR analyzed previously remain the same. Although the ``coronagraph'' may obscure a fraction of the IELR, it makes the detection of IELs much more robust.

\subsection{Implications and Future Work}
With the reddening quantities derived above, we recover the emission lines of OI 287 before dust extinction. Figure 10 displays the extinction corrected emission-line profiles of \lya\ and \civ, both normalized to the continuum. In both \lya\ and \civ, the unreddened BELs strongly outshines the IELs. The intensity ratio of IEL/BEL before extinction is only 2.8\% in \lya\ and 3.2\% in \civ, respectively. Such low ratios hereby elude detection of the IELs if BELs are not suppressed, which demonstrates the importance of obscuring the BELs for detecting the weak IELs.
 
As a comparison, we overplot the \lya\ and \civ\ emission-line profile of the composite spectrum (Vanden Berk et al. 2001) in Figure 10. The extinction corrected emission-line profiles of OI 287 are similar to those of the composite spectrum, indicating that the IELs found in OI 287 are common and in normal quasars the IELs are usually ``hidden'' in the spectrum.

Since there is nothing special about the IELs of OI 287, we infer that more objects similar to OI 287 could be found. Taking OI 287 as a prototype, we have found a few tens of analogues of OI 287 from the Baryon Oscillation Spectroscopic Survey (BOSS, Ahn et al. 2013). In addition, Alexandroff et al. (2013) recently presented a sample of candidate type II quasars selected to have strong lines of \lya\ and \civ\ with average FWHM $\sim$ 1500 \kmps. In follow-up observations, Greene et al. (2014) reported that some of these objects have a broad \ha\ component with FWHM up to 7500 \kmps. Their broadband SED and Balmer decrement indicate that these objects are moderately obscured. These characteristics are very similar to those of OI 287, implying that they are likely analogues of OI 287. The findings of these objects indicate that the IELR is common in AGNs and OI 287 may represent a population of AGNs. It is possible and meaningful to compile a large sample of objects similar to OI 287, which can be used to further study the properties of IELRs.
\section{SUMMARY}
With archived data and follow-up observations of the quasar OI 287, we presented a detailed analysis of its emission lines, broadband SED, and absorption lines. The emission lines are dominated by IELs with FWHM $\sim$ 2000 \kmps\ in the UV, since the corresponding BELs are heavily suppressed by dust obscuration as indicated by the Balmer decrement and intensity ratios of BELs in OI 287 to BELs in the composite quasar. The broad brand SED is identical to the composite quasar spectra in longer wavelengths, but clearly deviates from the composite quasar spectra short-ward portion, indicating the central accretion disk is also reddened. The $E(B-V)$ of reddened power law from the SED fitting is close to that estimated from the BELs. The absorption lines are consistent with partial obscuration: clouds fully cover the central accretion disk and BELR but do not block the outer regions, including the IELR, dusty torus, NELR and scattering region.

Based on these results, we discussed the properties of the IELR and the obscuring material. Assuming the IELR is virialized, we estimated its distance to the central black hole, $R_{\rm IELR} \sim$ 2.9 pc, which is similar to the dust sublimation radius of $R_{\rm sub} \sim$ 1.3 pc in OI 287. Comparison between photo-ionization model calculations and IEL measurements of \lya/\civ\ and EW(\lya) suggests that the IELR has a hydrogen density of $n_{\rm H} \sim 10^{8.8}-10^{9.4}~\rm cm^{-3}$, within the ranges often quoted for the dusty torus near the sublimation radius. Adopting such parameters and the covering factors estimated from the hot dust emission, we predicted the equivalent widths for all other permitted IELs, which are consistent with their observed values within the measurement uncertainties. These results provide another piece of evidence that the IELs originate from the inner region of the dusty torus through photo-ionization processes. The inferred location and physical properties strongly suggest that IELs originate from the inner part of the dusty torus. The fact that the central accretion disk and BELR are reddened by dust but the IELR and NELR are not implies that the obscuring material is likely located in the dusty torus. Associated with the dusty material, we identified the only one narrow absorption-line system in the spectrum of OI 287. Photo-ionization model calculations suggest that the obscuring material may originate from the dusty torus beyond the dust sublimation radius. Therefore, we speculate that both the IELR and obscuring material are part of the dusty torus. The IELR, located in the inner part of the dusty torus, is exposed to the central ionizing source and produces the IELs through photo-ionization processes; while the obscuring material in the outer part of the dusty torus obscures the central accretion disk and BELR as a ``coronagraph''.
\clearpage
We thank the anonymous referee for careful comments and helpful suggestions that led to the improvement of the paper. Many thanks to Sarah Bird for reading the manuscript and correcting the English writing. H.Z. thanks Jianmin Wang for the helpful discussion. This work is supported by the SOC program (CHINARE 2012-02-03), Natural Science Foundation of China grants (NSFC 11473025, 11033007, 11421303, 11503022, 11473305), National Basic Research Program of China (the 973 Program 2013CB834905), and Strategic Priority Research Program ``The Emergence of Cosmological Structures'' (XDB 09030200).

This research uses data obtained through the Telescope Access Program (TAP), which has been funded by the Strategic Priority Research Program ``The Emergence of Cosmological Structures'' (XDB 09000000), National Astronomical Observatories, Chinese Academy of Sciences, and Special Fund for Astronomy from the Ministry of Finance.

Some of the data presented in this paper were obtained from the Mikulski Archive for Space Telescopes (MAST). The Space Telescope Science Institute (STScI) is operated by the Association of Universities for Research in Astronomy, Inc., under the National Aeronautics and Space Administration (NASA) contract NAS5-26555. Support for non-HST MAST data is provided by the NASA Office of Space Science via grant NNX13AC07G and by other grants and contracts. The UKIDSS project is described in Lawrence et al. (2007). UKIDSS uses the UKIRT Wide Field Camera (WFCAM; Casali et al. (2007)) and a photometric system described in Hewett et al. (2006). The pipeline processing and science archive are described in Irwin et al. (2008) and Hambly et al. (2008). This publication makes use of data products from the WISE, which is a joint project of the University of California, Los Angeles, and the Jet Propulsion Laboratory/California Institute of Technology, funded by NASA.

Funding for SDSS and SDSS-II has been provided by the Alfred P. Sloan Foundation, Participating Institutions, National Science Foundation, U.S. Department of Energy, NASA, Japanese Monbukagakusho, Max Planck Society, and Higher Education Funding Council for England. The SDSS is http://www.sdss.org/.

SDSS is managed by the Astrophysical Research Consortium for the Participating Institutions. The Participating Institutions are the American Museum of Natural History, Astrophysical Institute Potsdam, University of Basel, University of Cambridge, Case Western Reserve University, University of Chicago, Drexel University, Fermilab, Institute for Advanced Study, Japan Participation Group, Johns Hopkins University, Joint Institute for Nuclear Astrophysics, Kavli Institute for Particle Astrophysics and Cosmology, Korean Scientist Group, Chinese Academy of Sciences (LAMOST), Los Alamos National Laboratory, Max-Planck-Institute for Astronomy (MPIA), Max-Planck-Institute for Astrophysics (MPA), New Mexico State University, Ohio State University, University of Pittsburgh, University of Portsmouth, Princeton University, United States Naval Observatory, and the University of Washington.
\clearpage
\begin{center}
APPENDIX
\end{center}
\begin{appendix}
\section{Improving IEL Measurements in the Partially Obscured Quasar with Suppressed BELs}
Since the UV BELs in the quasar OI 287 are suppressed, the IELs in the UV range become prominent and thus can be reliably measured. We made a simulation by measuring the \lya\ IEL as an example to investigate the dependence of the measurement accuracy of UV IELs over the IEL/BEL flux ratio. In addition, the measurement accuracy of the optical/NIR IELs can also be improved, because their shifts and profiles can be fixed to those of the prominent UV IELs. To inspect this fit strategy, we carried out another simulation by measuring the \ha\ IEL as an example.

We illustrate the simulation of measuring the \lya\ IEL in Figure A1. First, we construct the \lya\ emission-line spectrum by combining the three components of the NEL (Panel (a)), IEL (Panel (b)), and BEL (Panel (c)). Each component is generated by a single Gaussian and all of the emission-line parameters are set to those of OI 287, except that the BEL flux varies in the range of 500--250,000 $\times 10^{-17}~\rm erg~s^{-1}~cm^{-2}$ (corresponding to the IEL/BEL flux ratio in the range of 0.01--5). By adding random noise, the signal-to-noise of the newly constructed spectrum is set to that of OI 287. Then we fit the simulated spectrum using our automatic algorithm (Panel (d)) and finally obtain the best-fit output IEL (Panel (e)). We repeat the above procedure (both the spectrum construction and fitting) 500 times. With the input and output IELs, we investigate the measurement relative errors of IEL flux (defined as $\frac{flux({\rm output~IEL})-flux({\rm input~IEL})}{flux({\rm input~IEL})}$) as a function of the input IEL/BEL flux ratio (Panel (g)). The result clearly shows that the measurement errors are obviously reduced with increasing IEL/BEL flux ratio. Due to the suppression of \lya\ BEL, OI 287 has a large IEL/BEL flux ratio of $\approx$ 2.41 (red dashed line). However, if the \lya\ BEL of OI 287 was not suppressed, the value would be smaller to about 0.028 (blue dashed line, derived by making SMC extinction correction with $E(B-V) \approx 0.29$ mag, see Section 4.3 for details). The distributions of the measurement errors, when the \lya\ BEL is unsuppressed/suppressed, are shown in Panel (f) and Panel (h), respectively. By suppressing the \lya\ BEL of OI 287, the 1-$\sigma$ measurement errors are greatly reduced from 18.7\% to 0.8\%.

The simulation of measuring \ha\ IEL is shown in Figure A2. The process is similar to that of measuring the \lya\ IEL described above, but with some modifications. We construct the simulated spectrum by combining the NELs including \ha, \nii\ $\lambda\lambda$6548, 6583, and \sii\ $\lambda\lambda$6716, 6731 (Panel (a)), the IEL of \ha\ (Panel (b)) and the BEL of \ha\ (Panel (c)). All parameters are set to those of OI 287, except that the IEL flux varies in the range of 275--15,000 $\times 10^{-17}~\rm erg~s^{-1}~cm^{-2}$ (IEL/BEL flux ratio in the range of 0.01--1). The simulated spectrum is separately fitted in the following two cases: the redshift and profile of the input IEL are (1) free and (2) fixed to those of the \lya\ IEL. The result shows that the measurement errors in the two cases are both obviously reduced with increasing IEL/BEL flux ratio, but more effective in the second case (Panels (g) and (i)). Panels (f) and (h) display the distributions of the measurement errors in the two cases when the input IEL/BEL flux ratio equals that of OI 287 ($\approx 0.017$). By fixing the redshift and profile of the \ha\ IEL in OI 287, the 1-$\sigma$ measurement errors are reduced from 37.3\% to 15.4\%.
\end{appendix}

\clearpage
\begin{deluxetable}{ccccccc}
\tabletypesize{\small}
\tablewidth{0pt}
\tablecaption{Spectroscopic Data}
\tablehead{\colhead {Range} & {Slit}   & {$\lambda/\Delta\lambda$} & {Exp.Time} & {Instrument} & {Data} & {Reference} \\
                 \colhead {(\AA)}  & {($''$)} & {}                                          & {(s)}           & {}                  & {(UT)}  & {}}
\startdata
1590--2310              & 0.9 & 1300      & 1830 & \textit{HST}/FOS/G190H & 1992 May  21 & 1  \\
3800--9200              & 3.0 & 2000      & 3584 & SDSS                   & 2002 Oct. 31 & 2,3\\
9700--24600             & 1.1 & 3500      & 720  & P200/TripleSpec        & 2013 Feb. 23 & 4  \\
3150--5850; 7840--10700 & 2.0 & 800; 1700 & 1200 & P200/DoubleSpec        & 2014 Apr. 22 & 4  \\
\enddata
\tablerefs{(1) Kuraszkiewicz et al. 2002; (2) York et al. 2003; (3) Abazajian et al. 2009; (4) This work.}
\end{deluxetable}
\clearpage
\begin{deluxetable}{lcclc}
\tabletypesize{\normalsize}
\tablewidth{0pt}
\tablecaption{Photometric Data}
\tablehead{\colhead {Band} & {Value} & {Facility} & {Date} & {Reference} \\
                    {}     & {(mag)} & {}         & {(UT)} & {}          }
\startdata
FUV  & 20.855$\pm$0.194 & GALEX  & 2006 Feb. 15 & 1   \\
NUV  & 19.829$\pm$0.097 & GALEX  & 2006 Feb. 15 & 1   \\
$u$  & 19.009$\pm$0.024 & SDSS   & 2002 Jan. 14 & 2,3 \\
$g$  & 18.221$\pm$0.006 & SDSS   & 2002 Jan. 14 & 2,3 \\
$r$  & 17.680$\pm$0.006 & SDSS   & 2002 Jan. 14 & 2,3 \\
$i$  & 17.040$\pm$0.005 & SDSS   & 2002 Jan. 14 & 2,3 \\
$z$  & 16.607$\pm$0.011 & SDSS   & 2002 Jan. 14 & 2,3 \\
$Y$  & 15.903$\pm$0.005 & UKIDSS & 2009 Feb. 1  & 4   \\
$J$  & 15.450$\pm$0.004 & UKIDSS & 2007 Feb. 14 & 4   \\
$H$  & 14.679$\pm$0.005 & UKIDSS & 2009 May 2   & 4   \\
$K$  & 13.800$\pm$0.005 & UKIDSS & 2009 May 2   & 4   \\
$W1$ & 12.282$\pm$0.023 & WISE   & 2010 Apr. 10 & 5   \\
$W2$ & 11.200$\pm$0.022 & WISE   & 2010 Apr. 10 & 5   \\
$W3$ &  8.658$\pm$0.026 & WISE   & 2010 Apr. 10 & 5   \\
$W4$ &  5.972$\pm$0.048 & WISE   & 2010 Apr. 10 & 5   \\
\enddata
\tablerefs{(1) Morrissey et al. 2007; (2) York et al. 2000; (3) Abazajian et al. 2009; (4) Lawrence et al. 2007; (5) Wright et al. 2010.}
\end{deluxetable}
\clearpage
\begin{deluxetable}{l c cc cccccccccc}
\tabletypesize{\footnotesize}
\tablewidth{0pt}
\rotate
\tablecaption{Measurements of Emission Lines}
\tablehead{\colhead {Model} &{Component}&{Shift}  &{FWHM}   &\multicolumn{10}{c}{Flux} \\
                    {}      &{}         &{(\kmps)}&{(\kmps)}&\multicolumn{10}{c}{($\rm 10^{-17}~erg~s^{-1}~cm^{-2}$)} \\
                    \cline{5-14}
                    {}      &{}         &{}       &{}       &{\lya}&{\nv}&{\siiv}&{\civ}&{\mgii}&{\hr}&{\hb}&{\ha}&{\hei}&{\pr}}
\startdata
{       }& BELs&     -97$\pm      14$&    9117$\pm      38$&    1033$\pm      33$&$<     35$&$<     166$&     637$\pm      31$&    1951$\pm      26$&     991$\pm      30$&    3502$\pm      42$&   20285$\pm      80$&    2025$\pm     116$&
    1289$\pm     107$\\
{Model 1}& IELs&     -49$\pm       7$&    1974$\pm      22$&    2495$\pm      30$&$<     42$&$<     109$&     995$\pm     186$&      83$\pm      30$&      75$\pm      23$&     130$\pm      25$&     357$\pm      94$&     194$\pm      64$&
      36$\pm      23$\\
{       }& NELs&      56$\pm       1$&     641$\pm       4$&     449$\pm      23$&$<      6$&$<       9$&       --            &     275$\pm      18$&     150$\pm      11$&     313$\pm      12$&     958$\pm      43$&     318$\pm      34$&
     151$\pm       17$\\
\hline
{       }& BELs&     -85$\pm      12$&    8991$\pm      41$&     988$\pm      29$&$<     24$&$<     173$&     623$\pm      28$&    1945$\pm      29$&     989$\pm      27$&    3498$\pm      37$&   20288$\pm      73$&    2047$\pm     143$&
    1376$\pm     167$\\
{Model 2}& IELs&     -51$\pm       9$&    2025$\pm      26$&    2494$\pm      32$&$<     37$&$<     124$&    1005$\pm     190$&      72$\pm      28$&      72$\pm      21$&     113$\pm      18$&     271$\pm      85$&    205$\pm      72$&
      34$\pm      21$\\
{       }& NELs&      52$\pm       4$&     637$\pm      10$&     477$\pm      17$&$<      4$&$<       8$&       --            &     278$\pm      17$&     153$\pm      10$&     317$\pm      15$&     969$\pm      40$&     325$\pm      37$&
     147$\pm       22$\\
\enddata
\end{deluxetable}
\clearpage
\begin{figure}
\epsscale{0.8}
\plotone{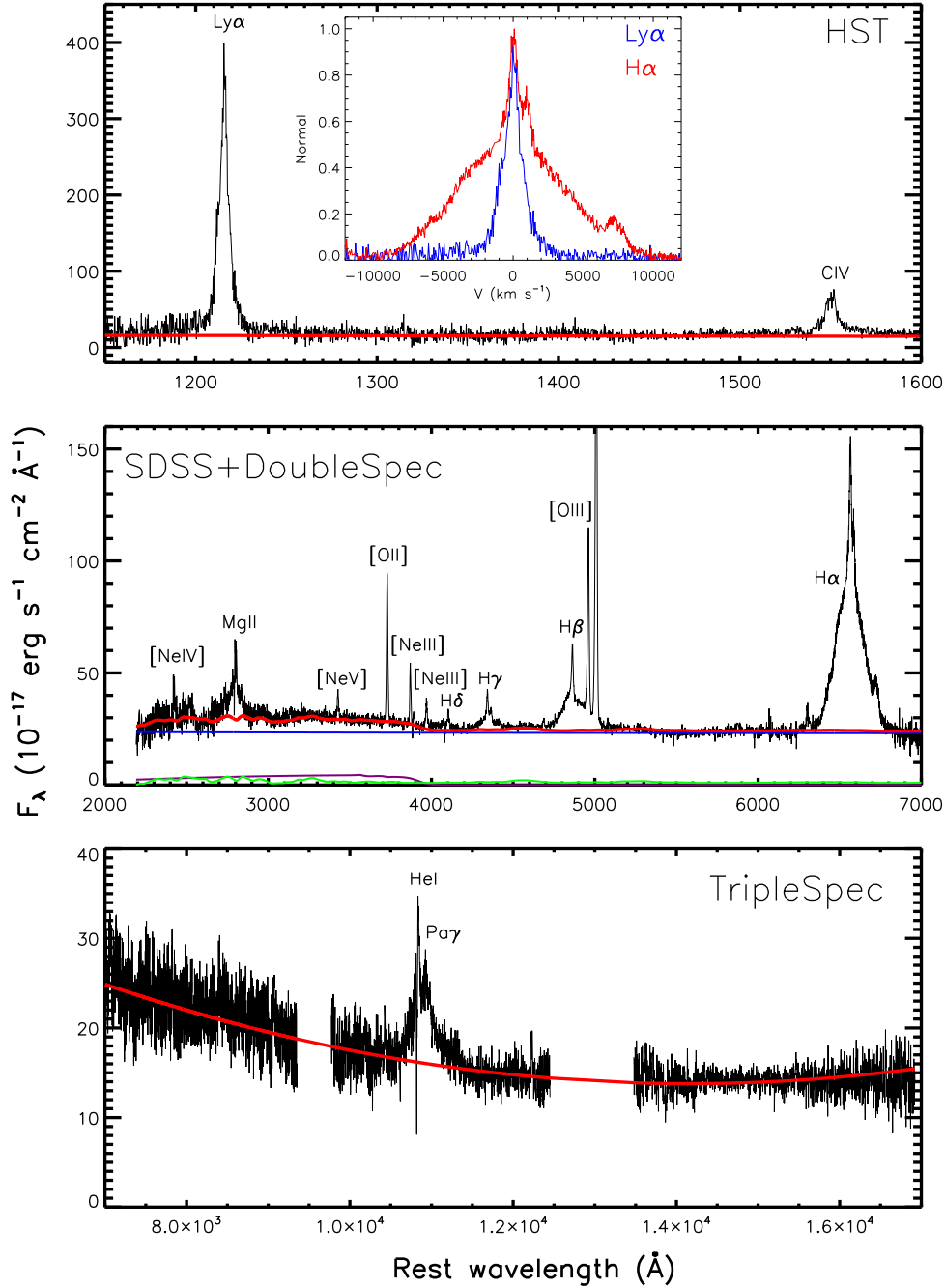}
\caption{Broadband spectra (black) of OI 287 overlaid with the continuum models (red). \textbf{Top}: UV spectrum taken by \textit{HST}/FOS and a power-law continuum model. \textbf{Middle}: Optical spectrum combined by the SDSS and DoubleSpec spectrum. The continuum model includes a power law (blue line), a Balmer continuum (purple line) and \feii\ pseudocontinuum (green line). \textbf{Bottom}: NIR spectrum obtained by TripleSpec and a second-order polynomial continuum model. Prominent emission lines are labelled in each panel. The UV emission lines of \lya\ and \civ\ are dominated by IELs, while BELs in \mgii, \hr, \hb, \ha, \hei\ and \pr\ are prominent in the optical and NIR spectrum. To make a clear comparison, we show the profiles of \lya\ and \ha\ in their common velocity space as an example in the insert panel.}
\end{figure}
\clearpage
\begin{figure}
\epsscale{0.5}
\plotone{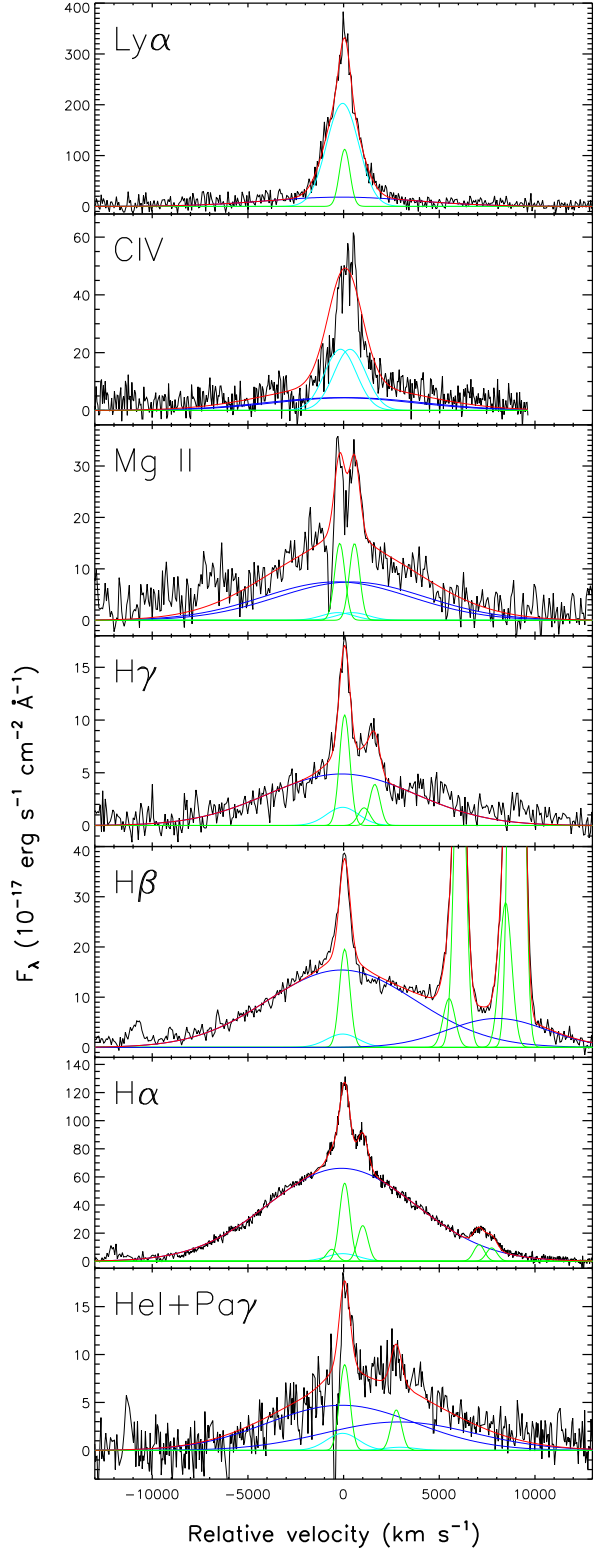}
\caption{Strong permitted emission lines of OI 287 shown in their common velocity space. From top to bottom, emission lines are sorted from shorter to longer wavelengths. Toward shorter wavelengths, BELs become weaker, while IELs become more prominent. We decompose these emission lines into the broad (blue), narrow (green), and intermediate-width (cyan) component.}
\end{figure}
\clearpage
\begin{figure}
\epsscale{1}
\plotone{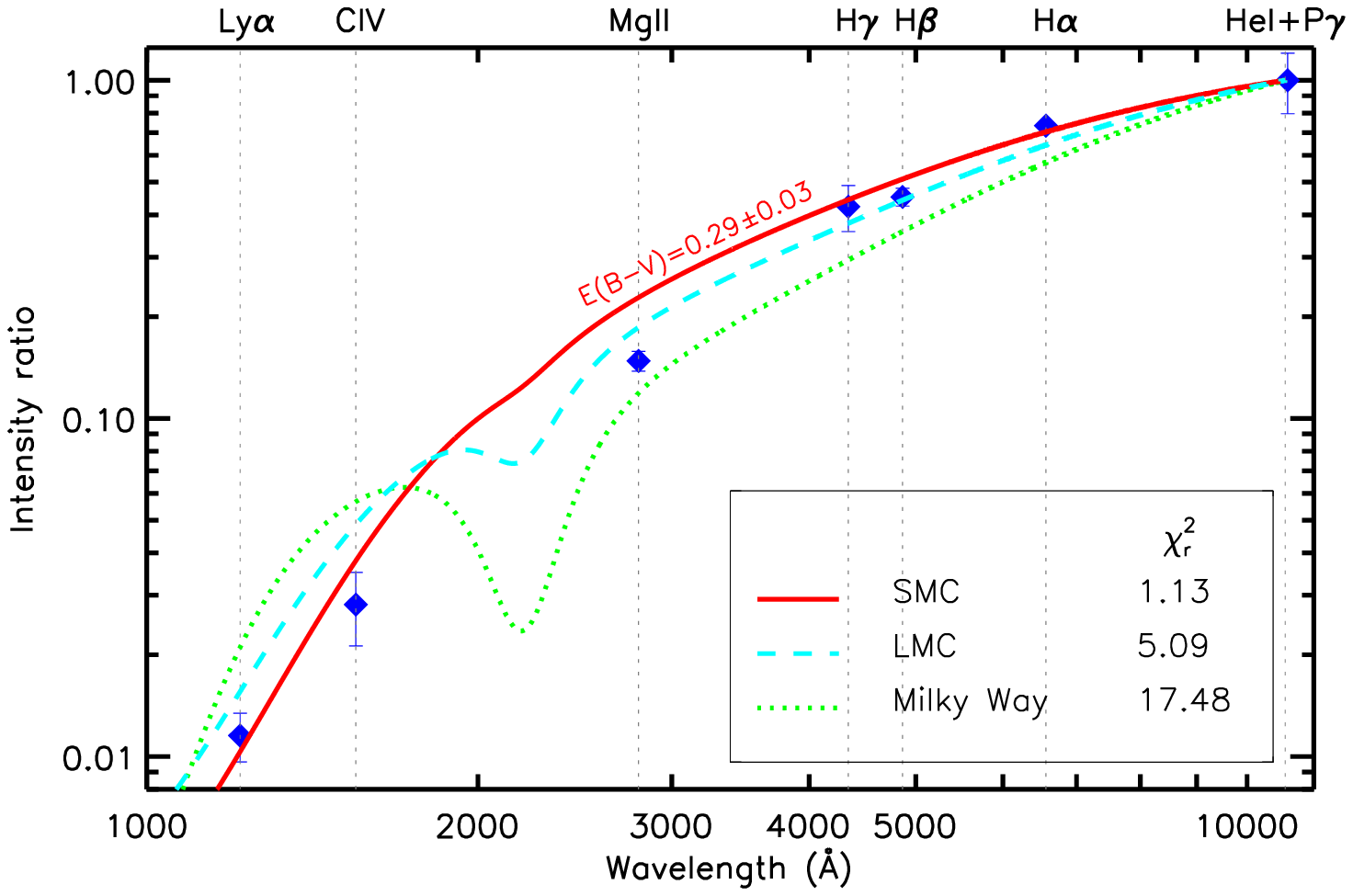}
\caption{Intensity ratios (blue diamond) of BELs in OI 287 to BELs in the composite quasar (\lya, \civ, \mgii, \hr, \hb, and \ha\ from Vanden Berk et al. (2001); \hei\ and \pr\ from Glikman et al. (2006)). The ratios are normalized to unity at \hei+\pr. From long to short wavelength lines, the intensity ratios gradually decrease, which suggests that the BELs are reddened. We fit the intensity ratios using three different extinction curves: SMC (red line), LMC (cyan dashed line) and Milky Way (green dotted line). All of the three extinction curve can describe the BEL intensity ratios, but SMC-like extinction ($E(B-V)$ = 0.29$\pm0.03$) is more likely compared with that of LMC and Milky Way.}
\end{figure}
\clearpage
\begin{figure}
\epsscale{1}
\plotone{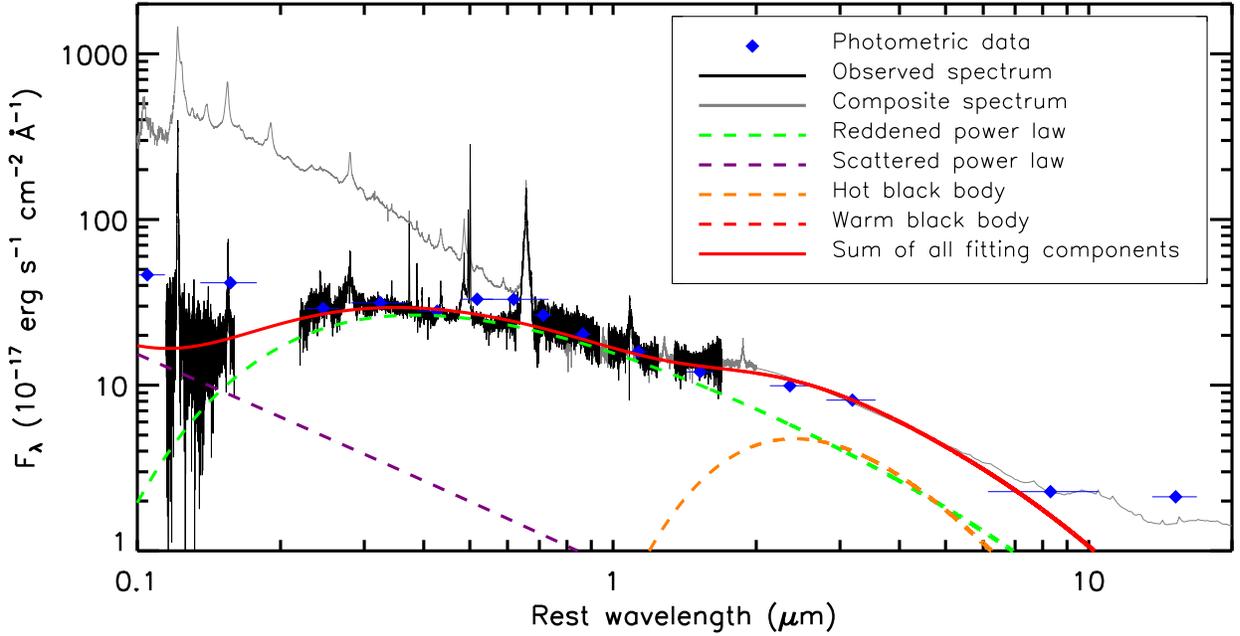}
\caption{Broadband SED of OI 287 in the rest frame from UV to NIR. We plot the observed spectra (black solid line) and the photometric data (blue diamond). The quasar composite spectrum (gray solid line) normalized at WISE--$W3$ is overplotted for comparison. In the long-ward portion $\lambda >$ 0.7~$\mu$m, the observed SED of OI 287 is nearly identical to the composite quasar spectrum, while gradually decreases in the short-ward portion $\lambda <$ 0.7~$\mu$m. We model the broadband SED using a reddened power law (green dashed line), a scattered power law (purple dashed line), and a hot (orange dashed line) and a warm (red dashed line) black body. The sum of all modelled components (red solid line) can roughly reproduce the continuum.}
\end{figure}
\clearpage
\begin{figure}
\epsscale{0.8}
\plotone{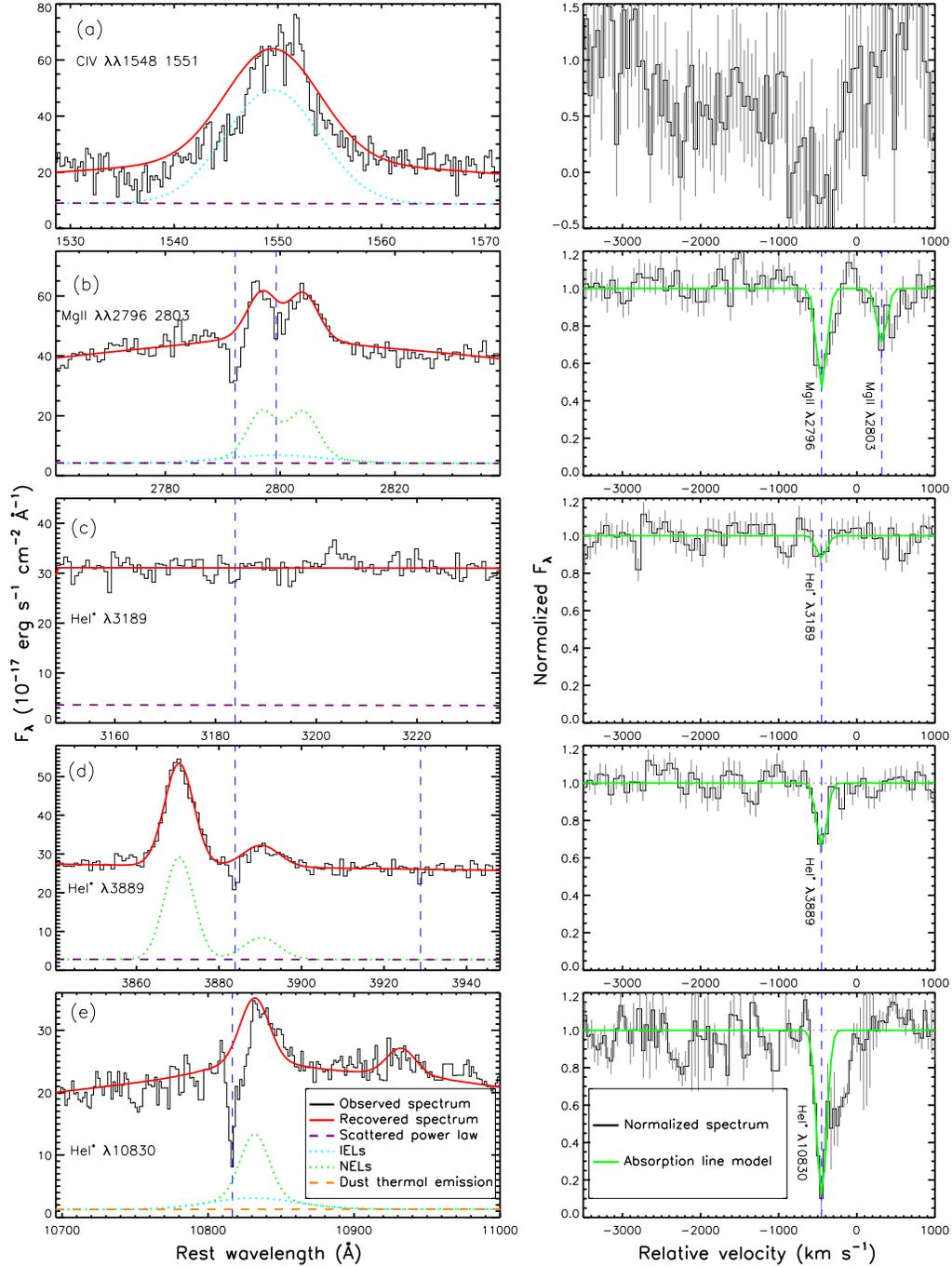}
\caption{\textbf{Left column}: Observed spectrum (black solid line) and recovered absorption-free spectrum (red solid line) in the vicinities of absorption lines including (a) \civ\ $\lambda\lambda$1548, 1551, (b) \mgii\ $\lambda\lambda$2796, 2803, (c) \hei$^*$ $\lambda$3189, (d) \hei$^*$ $\lambda$3889, and (e) \hei$^*$ $\lambda$10830. In each panel, we plot the components that are assumed to be not covered by the absorption gas, including the scattered power law (purple dashed line), IELs (cyan dotted line), NELs (green dotted line) and thermal dust emission (orange dashed line). \textbf{Right column}: Corresponding normalized absorption lines (black solid line). \civ\ absorption lines are not modelled due to their complex profiles and low S/N. Each of the \mgii\ doublet and \hei$^*$ multiplet is modelled using a single Gaussian (green solid line).}
\end{figure}
\clearpage
\begin{figure}
\epsscale{1}
\plotone{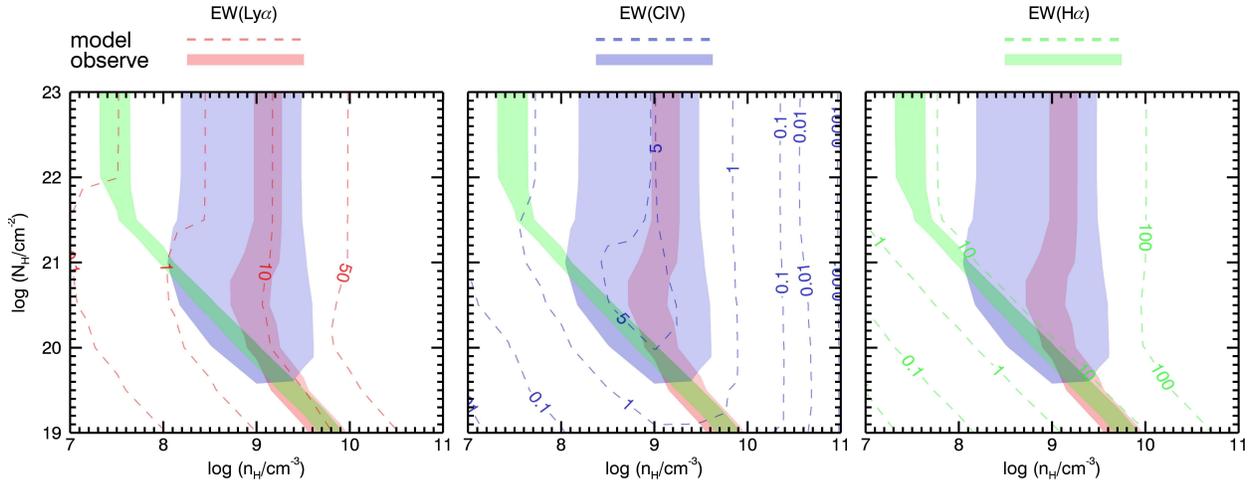}
\caption{Contours of EW(\lya) (red), EW(\civ) (blue) and EW(\ha) (green) as functions of $n_{\rm H}$ and $N_{\rm H}$ calculated by CLOUDY in the specific case: dusty gas with solar abundance, MF87 SED, $\lambda L_{\lambda}$(1450 \AA) $\approx 5.4 \times 10^{45}~\rm erg~s^{-1}$, $d=2.9$ pc (i.e., the hydrogen ionizing photon volume density $\Phi(\rm H)=10^{17.2}~\rm photons~s^{-1}~cm^{-2}$), and $f_{\rm C}=$ 7\%. In each panel, the dashed lines represent the baseline model and the filled areas represent the observed ranges for 1-$\sigma$ measurement errors. The overlapping region constrains the parameters of the IELR to a narrow range of $n_{\rm H} \sim 10^{8.8}-10^{9.4}~\rm cm^{-3}$ and $N_{\rm H} \sim 10^{19.6}-10^{20.2}~\rm cm^{-2}$.}
\end{figure}
\clearpage
\begin{figure}
\epsscale{1}
\plotone{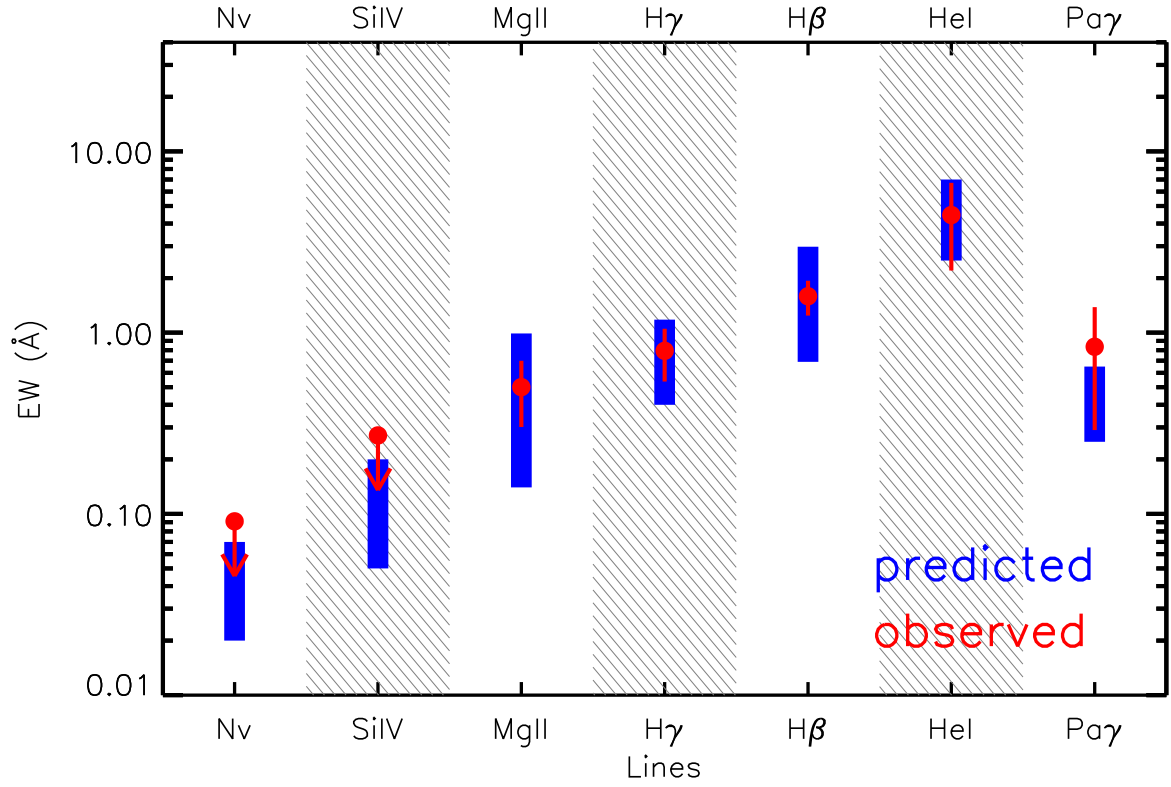}
\caption{Blue: IEL EWs predicted by the photo-ionization model. Red: IEL EWs observed in the spectrum of OI 287. The arrows denote the measured 3-$\sigma$ upper limits.}
\end{figure}
\clearpage
\begin{figure}
\epsscale{1.0}
\plotone{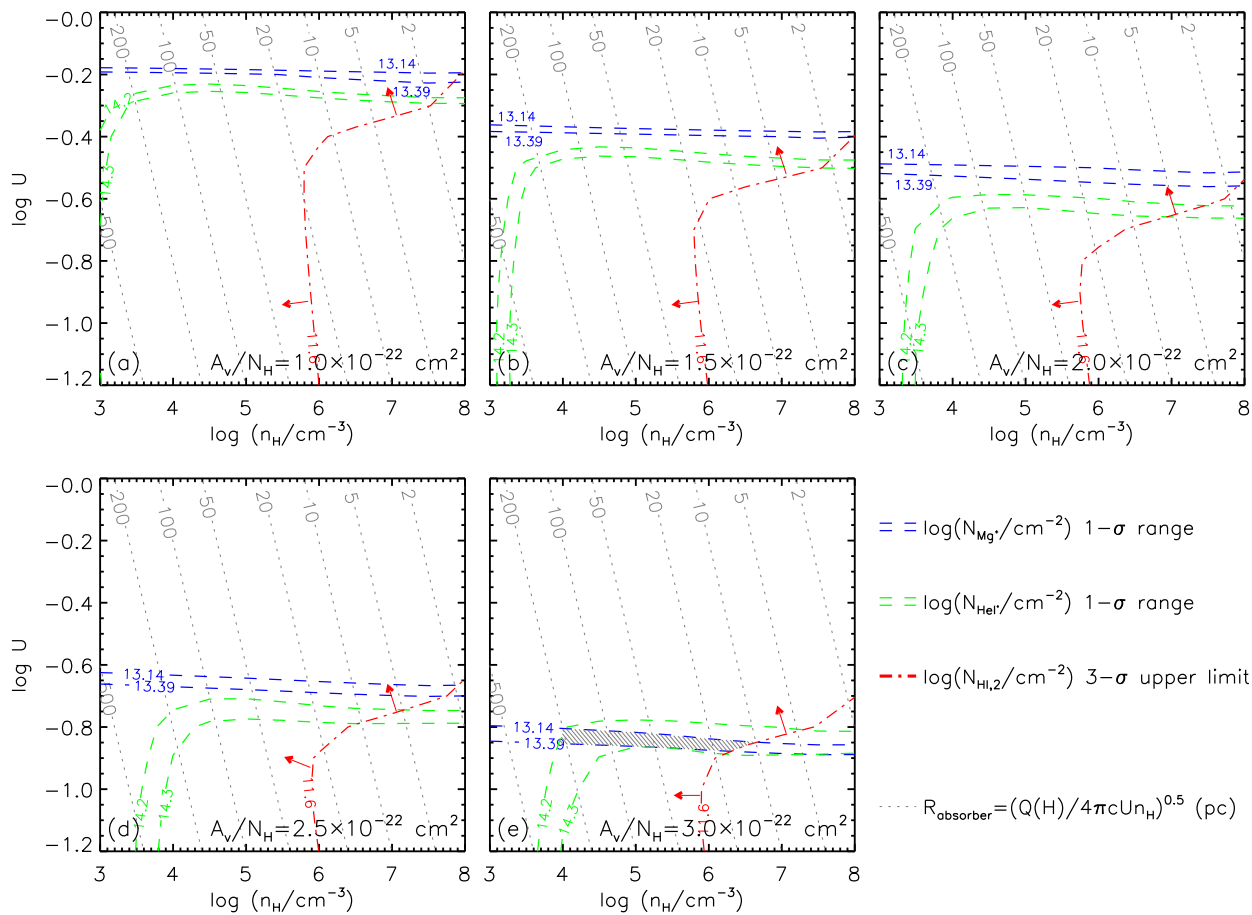}
\caption{Contours of $N_{\rm Mg^{+}}$, $N_{\rm HeI^*}$ and $N_{\rm HI,2}$ as functions of $n_{\rm H}$ and $U$ for dusty gas with increasing dust-to-gas ratio. The blue and green dashed lines denote the 1-$\sigma$ measurement error range of $N_{\rm Mg^{+}}$ and $N_{\rm HeI^*}$, respectively. The red dot-dashed lines represent the 3-$\sigma$ upper limit of $N_{\rm HI,2}$. When $A_{\rm v}/N_{\rm H}=3 \times 10^{-22}~\rm cm^{2}$ (Panel (e)), there is an overlap (filled area) of $U \sim 10^{-0.9}-10^{-0.8}$ and $n_{\rm H} \sim 10^{4.0}-10^{6.5} ~\rm cm^{-3}$. Within these parameter ranges, $R_{\rm absorber}$ is constrained in the range of $\sim$ 10--200 pc, as indicated by the gray dashed lines, which denotes the distance of the absorber to the central ionizing source, $R_{\rm absorber}=(Q({\rm H})/4\pi c U n_{\rm H})^{0.5}$.}
\end{figure}
\clearpage
\begin{figure}
\epsscale{1}
\plotone{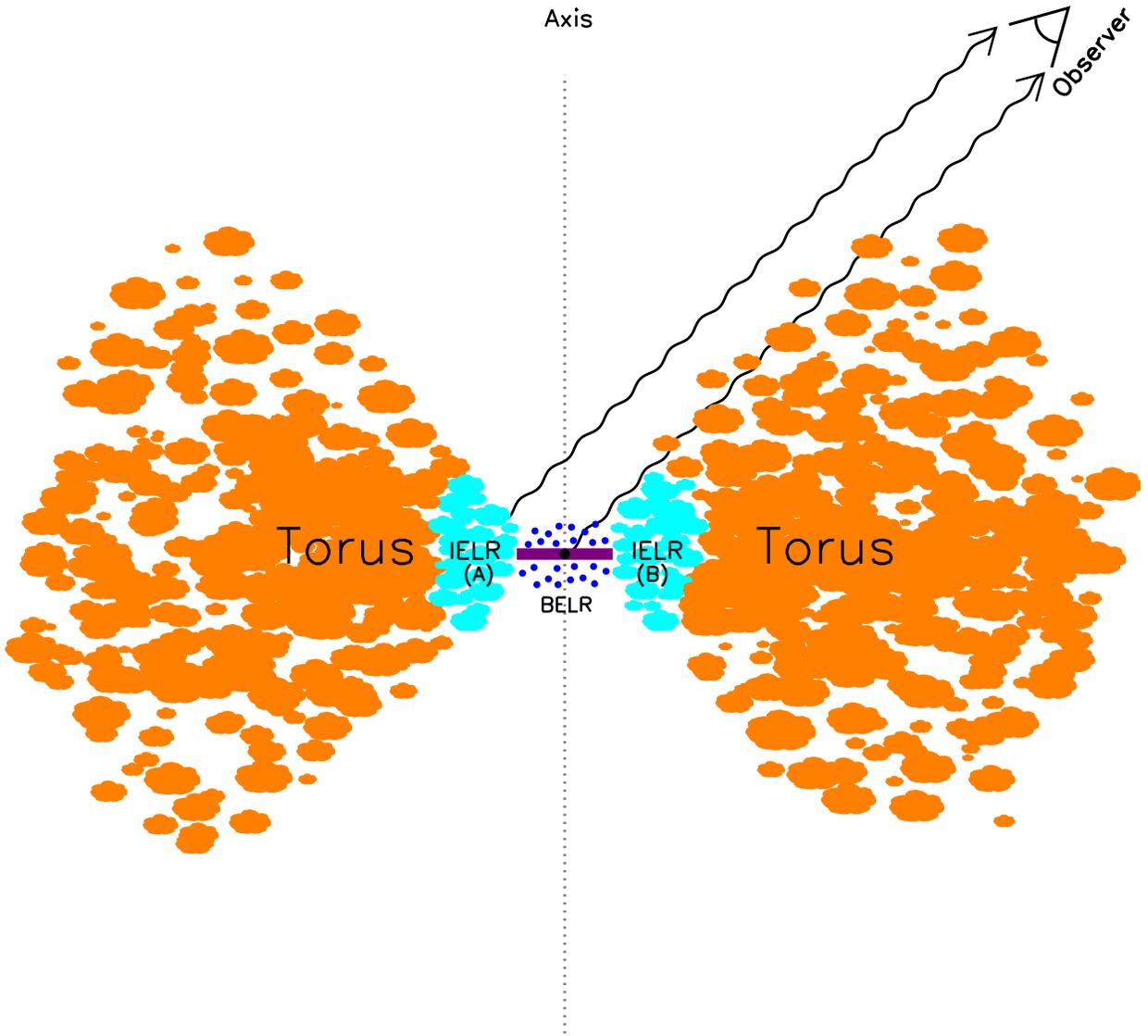}
\caption{Cartoon of detecting the IELR with the dusty torus as a ``coronagraph''. The lines-of-sight to the central accretion disk and the BELR are obscured by the dusty material near the boundary of the dusty torus, which results in the observed reddened SED and BELs. However, the line-of-sight to the IELR is not fully obscured. Some fraction of the IELR (e.g., the far side of the IELR, marked as ``A'') can be directly observed, which yields the observed prominent UV IELs, although some fraction of the IELR (e.g., the near side of the IELR, marked as ``B'') cannot be observed.}
\end{figure}
\clearpage
\begin{figure}
\epsscale{1}
\plotone{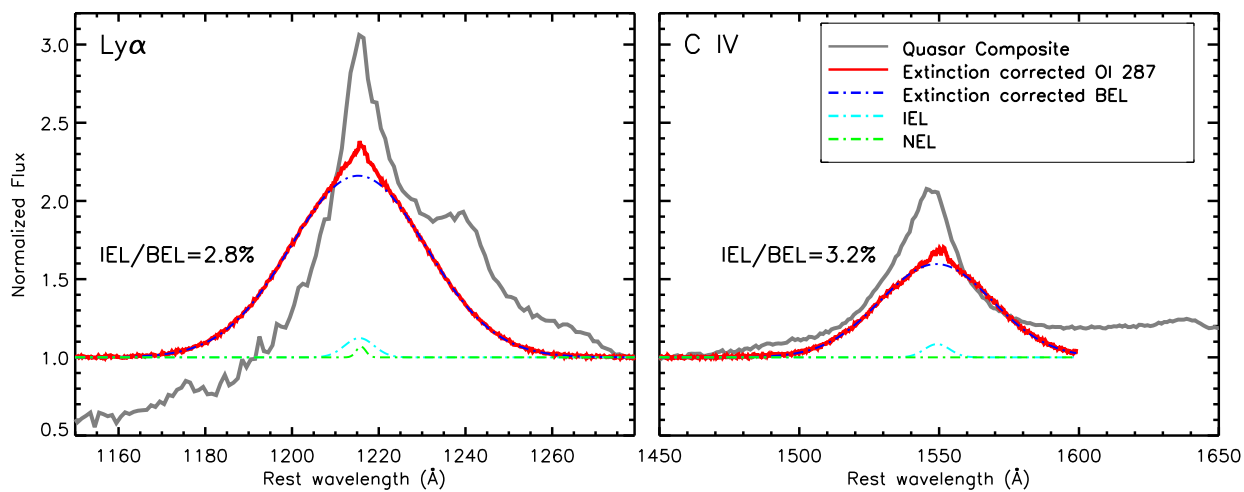}
\caption{\textbf{Left:} Extinction corrected \lya\ profile of OI 287 (red) and quasar composite (Vanden Berk et al. 2001; gray). Both emission lines are normalized to the continuum. We plot the NEL (green), IEL (cyan), and extinction corrected BEL (blue). \textbf{Right:} Same as the left panel but for \civ. The intensity ratios of IEL/BEL in OI 287 before extinction are only 2.8\% in \lya\ and 3.2\% in \civ, respectively.}
\end{figure}
\clearpage
\figurenum{A1}
\begin{figure}
\epsscale{0.9}
\plotone{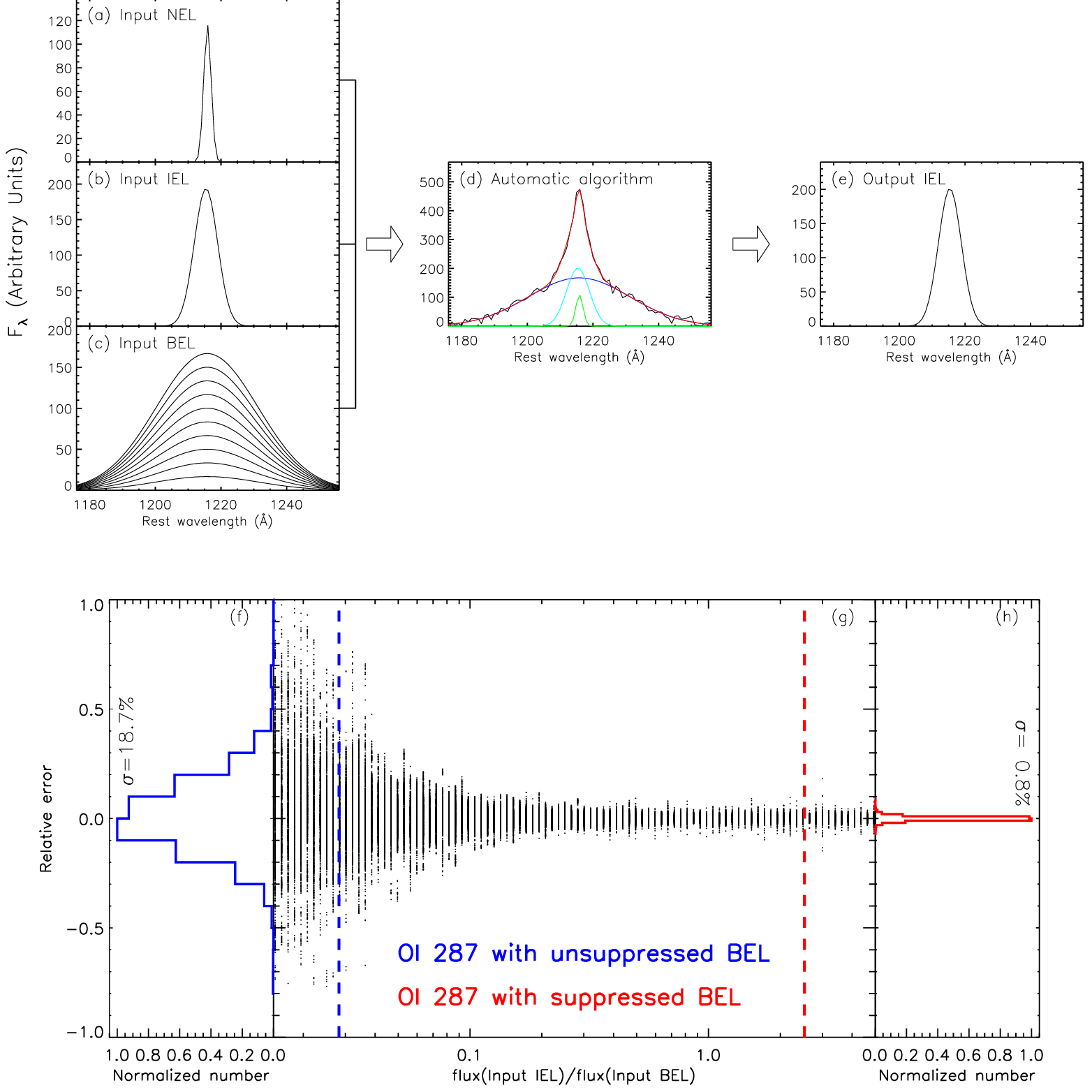}
\caption{Demonstration of improving \lya\ IEL measurement accuracy with increasing IEL/BEL flux ratio. \textbf{Panels (a--c)}: Input three components of the NEL, IEL and BEL. All of their parameters are set to those of OI 287, except that the BEL flux varies in the range of 500--250,000 $\rm \times 10^{-17}~erg~s^{-1}~cm^{-2}$ (corresponding to the IEL/BEL flux ratio in the range of 0.01--5). \textbf{Panel (d)}: Simulated spectrum (black) is decomposed into the NEL (green), IEL (cyan), and BEL (blue) through our automatic algorithm. The red line denotes the sum of all these best-fit components. \textbf{Panel (e)}: Best-fit output IEL. \textbf{Panel (g)}: Relative measurement errors of \lya\ IEL flux (defined as $\frac{flux({\rm output~IEL})-flux({\rm input~IEL})}{flux({\rm input~IEL})}$) as a function of the input IEL/BEL flux ratio. The blue and red dashed lines denote IEL/BEL flux ratios when the \lya\ BEL of OI 287 is unsuppressed and suppressed. \textbf{Panels (f) and (h)}: Distributions of the measurement errors for OI 287 in the two cases, with 1-$\sigma$ measurement errors greatly reduced from 18.7\% to 0.8\%.}
\end{figure}
\clearpage
\figurenum{A2}
\begin{figure}
\epsscale{1}
\plotone{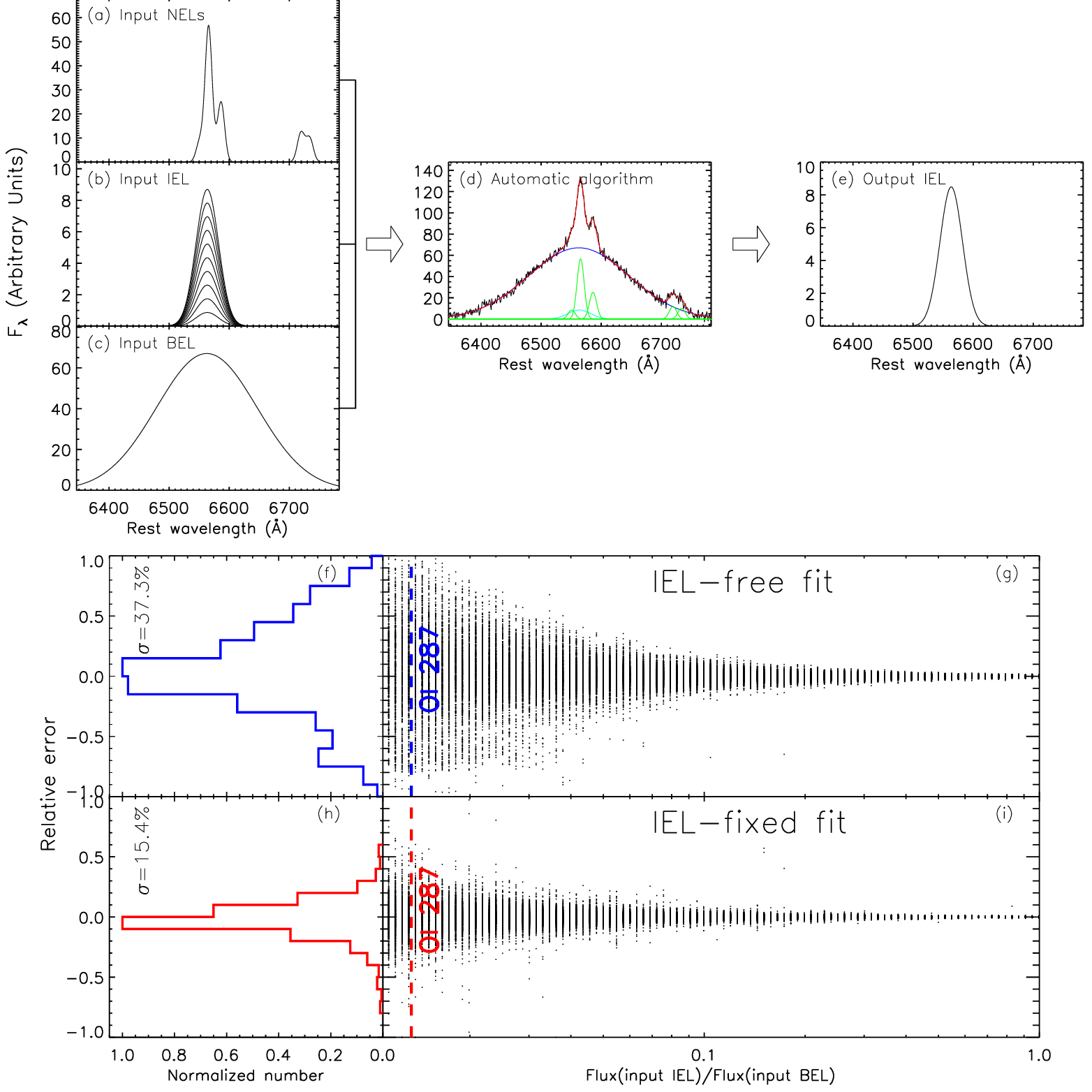}
\caption{Demonstration of improving \ha\ IEL measurement by fixing its redshift and profile to those of \lya\ IEL. \textbf{Panels (a--c)}: Input three components of NELs including \ha, \nii\ $\lambda\lambda$6548, 6583, and \sii\ $\lambda\lambda$6716, 6731, IEL of \ha, and BEL of \ha. All parameters are set to those of OI 287, except that the IEL flux varies in the range of 275--15,000 $\times 10^{-17}~\rm erg~s^{-1}~cm^{-2}$ (IEL/BEL flux ratio in the range of 0.01--1). \textbf{Panel (d)}: Simulated spectrum (black) is decomposed into the NELs (green), IEL (cyan), and BEL (blue). \textbf{Panel (e)}: Best-fit output IEL. \textbf{Panels (g) and (i)}: Relative measurement errors of IEL flux as a function of the input IEL/BEL flux ratio when the redshift and profile of the input IEL are free and fixed to those of \lya\ IEL. The dashed lines denote the IEL/BEL flux ratio of OI 287. \textbf{Panels (f) and (h)}: Distributions of the relative errors for OI 287 in the two cases, with 1-$\sigma$ relative error reduced from 37.3\% to 15.4\%.}
\end{figure}
\end{document}